# Validity of Fluctuation Theorem on Self-Propelling Particles

Ryo Suzuki, Hong-Ren Jiang, and Masaki Sano

*Department of Physics, The University of Tokyo,*
*Hongo 7-3-1, Bunkyo-ku, Tokyo 113-0033, Japan*
(Dated: April 29, 2011)

The experimental application of fluctuation theorem (FT) on nanometer to submicrometer sized systems has received thorough attention in the past several years. Nonetheless, the employment of FT on self-propelling objects has seldom been performed due to experimental difficulties. In this paper, we applied FT on a doublet that is comprised of asymmetrically coated self-propelling colloidal particles (Janus particles), which experiences a continuous rotation by applying AC electric field. We found that FT is valid for rotating self-propelling colloids by evaluating the estimated rotary torque to coincide with another form of torque estimation, which considers the rotation speed of the doublet and a simple Stokes equation.

Recent developments in the manipulation and observation of small objects that are of the order of nanometer to submicrometer in size, have provided many possible ways into investigating the non-equilibrium nature of fluctuations in a system. As a result, the application of fluctuation theories of non-equilibrium statistical mechanics [1-6] to such small systems have been performed experimentally, ranging from colloidal particle systems [7-9] to biological systems [10-13], giving quantitative insights into systems far from equilibrium.

Alongside the interests on fluctuation theorem (FT) and its verification and applications to small systems were fluctuation dominates, the properties of self-propelling microscopic objects and its motion [14-17] have received considerable attention, for they are fundamental problems in nonequilibrium physics. Consequently, the prospect of connecting FT and self-propelling objects is captivating and is at the same time indispensable, owing to the limited information and experiment that can be obtained and conducted for self-propelling objects at microscopic scales. Nevertheless, the validity of FT on self-propelling non-equilibrium systems has scarcely been tested. In this paper, we experimentally study a spinning doublet that is comprised of asymmetrically coated self-propelling colloidal particles and evaluate the validity of FT on the system, on the same basis as that of Ref. [13].

*Experimental system* --- Colloidal particles that have two sides with different features are called Janus particles. Various types of Janus particles can be devised and synthesized, and although their application is relatively new, it has become a rapidly expanding research field [18]. Henceforth, Janus particles will represent metallodielectric particles, which was used in the experiments. Such Janus particles are known to demonstrate two-dimensional self-propulsive motion in AC field [19], which exhibit a plethora of interesting phenomena since the motion is not restricted (compare for example Ref. [20]). Not only is this two-dimensional motion novel but also it is both appropriate and beneficial when experimentally studying self-propelling microscopic systems since no information on the particle motions is lost. Self-propulsive motion of Janus particles is understood to be governed by ICEO (Induced-charge electro-osmosis) and originated by ICEP (Induced-charge electrophoresis), and is perpendicular to the electric field [20-22]. Here, we used the same experimental system as the one mentioned in Ref. [19], where Janus particles in pure water is placed inside a thin chamber that is constructed by sandwiching two ITO glasses together with a spacer of thickness approximately 40μm [Fig. 1(a)]. Janus particles are prepared by the method introduced in Ref. [23], coating one side of polystyrene particles (Cat17134: Polysciences) with approximately 25nm of gold. The two-dimensional motion of Janus particles are enhanced by increasing applied electric field, as is shown in Fig. 1(b).

Since the Janus particles are subjected to AC field, this system has two distinctive frequency regions concerning both the direction of self-propulsive motion and the interaction between particles. For pure water, at lower AC frequencies (<30kHz), Janus particles self-propel in the direction of the polystyrene side, as mentioned previously, and the interaction has a repulsive nature [Fig. 1(c)]. On the contrary, for higher AC frequencies (>30kHz), the direction of self-propulsion is opposite to that of the lower AC frequency case, also having attractive interactions between particles, which is again the reverse of the interaction found in the other frequency region [Fig. 1(c)]. Changing the frequency and fixing the applied voltage at 1.5V, the velocity of Janus particles for pure water condition switches at approximately 30kHz [Fig. 1(e)], where positive values indicate self-propulsion in the direction of the polystyrene half and vice-versa. Moreover, this characteristic switching frequency changes for different NaCl concentrations, as is also shown in Fig. 1(e). We conducted our experiments in the higher AC frequency region, where spinning Janus doublets [Fig. 1(f)] could be found. Like is depicted in Fig. 1(d), one particle is affixed to the bottom ITO glass and due to the nature of Janus particles at the high AC frequency region, another particle binds to the static (in the sense that the particle does not change its position, however, it does rotate on the spot) one to create a Janus doublet. In this frequency region, the adhesion of particles to the bottom ITO glass is fortuitous and since we do not apply large electric fields, few particles stick and initiates the creation of Janus doublets. The outer particle attached to the central static particle will exhibit circular motion. Here, the rotation is either clockwise or counterclockwise, depending on the orientation of the metal-coated side, which is purely indeterministic.

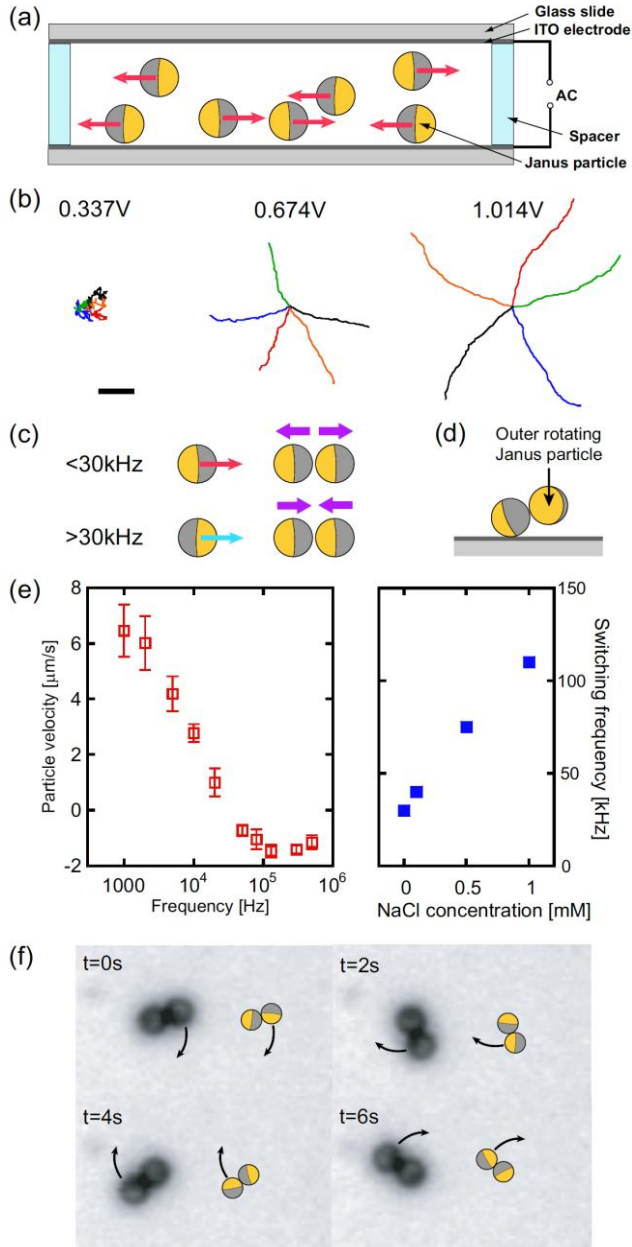

FIG. 1: (a) Schematic figure of the experimental setup (side view). Two ITO electrodes face each other, separated by a spacer of thickness approximately 40μm. An AC electric field is applied to the system, causing the Janus particles to move. Yellow half represents the metal-coated side. Not to scale. (b) Trajectories of 5 particles over 5 sec for 0.337V, 0.674V, and 1.014V at 5kHz. Scale bar: 3μm. (c) Direction of motion and particle interaction for lower (<30kHz) and higher (>30kHz) AC frequencies. At higher frequencies, direction of motion switches and particle interaction becomes attractive. (d) Schematic figure of a Janus doublet (side view). (e) Particle velocity for various AC frequencies and swithing frequency for different NaCl salt concentration. (f) Snapshot of a Janus doublet taken from the top, for $t = 0, 2, 4,$ and $6s$. The arrows depict the direction of the Janus doublet rotation. A drawing of the corresponding rotating doublet is shown on the right for clarity.

*Rotating Janus doublet* --- In our experiments, we used 3μm Janus particles in pure water and applied AC electric field at 400kHz, which was the optimal condition for the doublets to appear while preventing severe adhesion of the particles to the bottom ITO glass. The images of the rotation of Janus doublets were captured by a CCD camera connected to a microscope, at 10 frames per second for several minutes. The rotational angle of the Janus doublet, $\theta(t)$, was obtained from the images and one part of the time series is plotted in Fig. 2(a) for applied voltage 0.922V (red) and 1.861V (blue), with an example of the x-y position of the spinning outer Janus particle (at 0.922V) in the inset. Time series of the rotating angle $\theta(t)$ for all conditions are both adequately continuous and fluctuating. Hence, the application of FT the rotating Janus doublet system is appropriate. Moreover, rotational speed of Janus doublet was obtained by a linear fit of the time series. Figure 2(b) shows the resulting rotational speed. Increase in the rotational speed was observed by increasing the applied voltage to the system, having a proportional relation to the applied voltage squared. This fact coincides with the self-propelling velocity of Janus particles described in previous studies [20-22] as $\propto \varepsilon a E^2/\eta$, where $a$ is the particle radius, $E$ is the applied electric field, $\varepsilon$ and $\eta$ are the dielectric constant and viscosity of the bulk solvent, respectively.

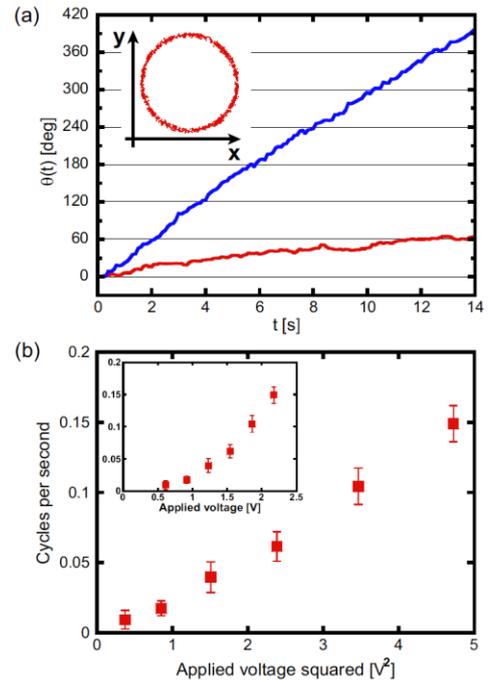

FIG. 2: (a) Time series of the rotational angle $\theta(t)$ of the Janus particle, for applied voltage 0.922V (red) and 1.861V (blue). One part of the time series is shown here. The continuous and fluctuating nature of the rotation is demonstrated. Inset: An example of the x-y position of the rotating outer Janus particle (for condition 0.922V). (b) Rotational speed of the Janus doublet, showing a proportional relation to the applied voltage squared. Inset: Original rotational speed.

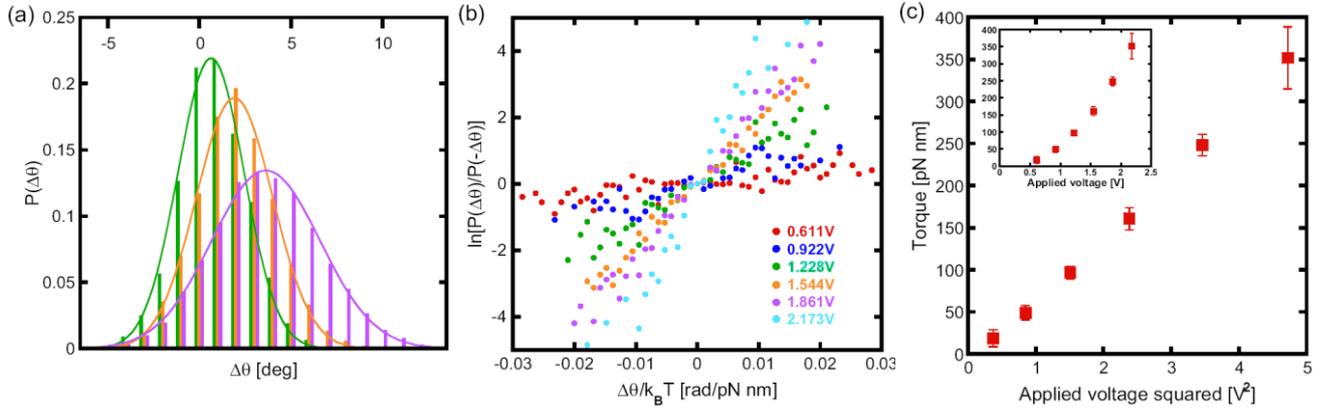

FIG. 3: (a) Histograms of the probability distribution of $\Delta\theta$ for 1.228V (green), 1.544V (orange), and 1.861V (purple). With increase in applied voltage, the probability distribution broadens and shifts to a larger value of $\Delta\theta$. Also, negative values for $\Delta\theta$ can be seen, demonstrating the possibility of counter rotation due to fluctuation. Three conditions are shown for clarity. Solid lines: Gaussian of data. (b) $\ln[P(\Delta\theta)/P(-\Delta\theta)]$ as a function of $\Delta\theta/k_BT$, Eq. 2, for all conditions, 0.611V (red), 0.922V (blue), 1.228V (green), 1.544V (orange), 1.861V (purple), and 2.173V (light blue). Inclination of the slope increases for larger applied voltage, implying larger $\tau$ with increase in applied voltage to the system. (c) Rotary torque of Janus doublets estimated from the inclination in (b). Torque is also proportional to the applied voltage squared, coinciding with the tendency seen for the rotational speed [Fig. 2(b)]. Inset: Original rotary torque.

*Fluctuation theorem* --- As in Ref. [13], we estimated the rotary torque acting on the outer rotating Janus particle via FT. Here, assuming that the rotation of the Janus doublet is continuous, the time evolution of $\theta(t)$ can be written by a Langevin equation as

$$\mu \frac{d\theta}{dt} = \tau + \xi(t).$$

The above $\mu$ and $\tau$ correspond to the frictional drag coefficient of the rotating particle and the rotary torque acting on the doublet. The rotary torque has the relation $\tau = \mu\omega$, where $\omega$ is the mean angular velocity of the doublet. Also, $\xi$ denotes random force which is an effect of thermal noise, where $\langle\xi(t)\xi(t')\rangle = 2\mu k_B T \delta(t-t')$. $k_B$ here is the Boltzmann constant and $T$ is room temperature. Presuming that $\tau$ is constant, Eq. 1 can be expressed in the context of FT for the rotary torque measurement [13] as

$$\ln\left[\frac{P(\Delta\theta)}{P(-\Delta\theta)}\right] = \tau \frac{\Delta\theta}{k_B T}.$$

Here, $\Delta\theta = \theta(t+\Delta t) - \theta(t)$ and $P(\Delta\theta)$ represents the probability distribution of $\Delta\theta$. Consequently, in measuring the rotation of the Janus doublet and attaining information on $\theta(t)$, one can indirectly obtain the rotary torque within a system where direct measurements are difficult in many aspects, since the particles are self-propelling.

In Fig. 3(a), $P(\Delta\theta)$ of Janus doublet rotation for 1.228V (green), 1.544V(orange), and 1.861V(purple) are shown as histograms, where $\Delta t = 0.1$s. Here, only three conditions are plotted for convenience, in order to clearly illustrate the changes that can be observed in the probability distributions by varying the applied voltage. The probability distribution of $\Delta\theta$ demonstrate negative values, which exemplifies rotation counter to the main rotating direction. For larger voltage, namely faster spinning of the doublet, the width of the probability distribution increases and shifts to a larger value of $\Delta\theta$, which is a manifest result. Moreover, the relation in Eq. 2 is plotted, for all applied voltage conditions, in Fig. 3(b). Here, the inclination of the slope, indicating the comparison between $\ln[P(\Delta\theta)/P(-\Delta\theta)]$ and $\Delta\theta/k_BT$, increases with increase in applied voltage, implying that $\tau$ increases for larger applied voltage. Torque acting on the Janus doublet is then estimated by a linear fit of the slope in Fig. 3(b) and the result is shown in Fig. 3(c), where $\tau$ is proportional to the applied voltage squared. This tendency coincides with the rotational speed [Fig. 2(b)].

*Validity* --- To appreciate the torque estimation via FT using rotational angle measurements, we compared the result with another torque estimation using the rotational speed and Stokes' drag force. Henceforth, $\tau_{FT}$ and $\tau_S$ will represent the torque estimated from FT and Stokes equation, respectively. We estimated $\tau_S$

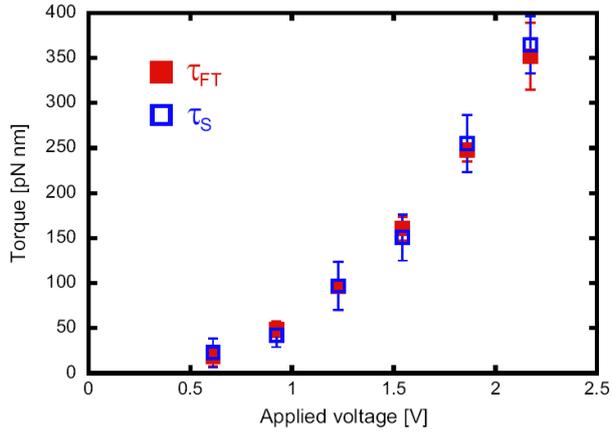

FIG. 4: Comparison between $\tau_{FT}$(red) and $\tau_S$. A good agreement can be seen between the two estimated rotary torques. Thus, implying the validity of FT on self-propelling induced rotating Janus doublet.

as
$$\tau_S = Fd, \qquad F = 6\pi\eta RV\beta,$$

where $\eta$ is the viscosity of the bulk solvent, $R$ is the radius of doublet, $V$ is the doublet velocity, and $d$ is the distance between the two particles that constitute the doublet, which is in principle 3μm. Here, the drag must be corrected for the proximity of the rotating Janus particle to the bottom surface by introducing $\beta$, which according to Faxen's law [24] is

$$\beta^{-1} = 1 - \frac{9}{16}\left(\frac{a}{h}\right) + \frac{1}{8}\left(\frac{a}{h}\right)^3 - \frac{45}{256}\left(\frac{a}{h}\right)^4 - \frac{1}{16}\left(\frac{a}{h}\right)^5,$$

where $h$ is the height of the particle center from the bottom surface. From the images of the rotating Janus doublets, for all applied voltage conditions, the average value of $h$ was $2.45 \pm 0.02$μm. This value was employed for the calculation of $\beta$, resulting in $\beta = 1.53$. Furthermore, for the particle velocity $V$ in Eq. 3, the rotational speed mentioned earlier in Fig. 2(b) was used. Figure 4 compares $\tau_S$ to $\tau_{FT}$, evincing the validity of the FT method in estimating the torque acting on the rotating Janus particle.

*Conclusion* --- We experimentally studied the application of FT on self-propelling Janus particles, by means of estimating the rotary torque acting on a Janus doublet. We found that this torque ($\tau_{FT}$) increased with increase in the voltage applied to the experimental system, having a proportional relation to the applied voltage squared. The validity of this result was evaluated by comparing $\tau_{FT}$ to a torque $\tau_S$ estimated via a different method, in which the rotating speed was used together with Stokes equation. A consistency in the two torques, $\tau_{FT}$ and $\tau_S$, elucidates the validity of the FT method. Such approach is influential, especially when dealing with self-propelling objects where direct measurements are indubitably problematic and details of the object itself are not fully understood or absent, which is usually the case. Further, using the Janus doublets, experimental investigation of the fluctuation-dissipation theorem for active colloidal systems can be anticipated [25]. The richness of interesting phenomena seen in our experimental system, along with the application of FT, merits for further attention and research, and may open new doors into the investigation of nonequilibrium physics.

We would like to acknowledge the members of our laboratory for fruitful discussions. This work is supported by Grant-in-Aid for Scientific Research (21244061) from MEXT Japan.